\documentclass[twocolumn,linenumbers]{aastex631}

\begin{document}

\title{Tilted circumbinary planetary systems as efficient progenitors of free-floating planets}

\author[0000-0002-4489-3491]{Cheng Chen}
\affiliation{School of Physics and Astronomy, University of Leeds\\
Sir William Henry Bragg Building, Woodhouse Ln., Leeds LS2 9JT, UK}

\author[0000-0003-2401-7168]{Rebecca G. Martin}
\affiliation{Department of Physics and Astronomy,  University of Nevada, Las Vegas\\
4505 South Maryland Parkway, Las Vegas, NV 89154, USA}
\affiliation{Nevada Center for Astrophysics, University of Nevada, Las Vegas\\
4505 South Maryland Parkway, Las Vegas, NV 89154, USA}

\author[0000-0002-4636-7348]{Stephen H. Lubow}
\affiliation{Space Telescope Science Institute \\
3700 San Martin Drive, Baltimore, MD 21218, USA}

\author[0000-0002-2137-4146]{C.J. Nixon}
\affiliation{School of Physics and Astronomy, University of Leeds\\
Sir William Henry Bragg Building, Woodhouse Ln., Leeds LS2 9JT, UK}

\begin{abstract}

The dominant mechanism for generating free-floating planets has so far remained elusive. One suggested mechanism is that planets are ejected from planetary systems due to planet-planet interactions. However, instability around a single star requires a very compactly spaced planetary system. We find that around binary star systems instability can occur even with widely separated planets that are on tilted orbits relative to the binary orbit due to combined effects of planet-binary and planet-planet interactions,  especially if the binary is on an eccentric orbit. We investigate the orbital stability of planetary systems with various planet masses and architectures. We find that the stability of the system depends upon the mass of the highest mass planet. The order of the planets in the system does not significantly affect stability but, generally, the most massive planet remains stable and the lower mass planets are ejected. The minimum planet mass required to trigger the instability is about that of Neptune for a circular orbit binary and a super-Earth of about $10$ Earth masses for highly eccentric binaries. Hence, we suggest that planet formation around misaligned binaries can be an efficient formation mechanism for free-floating planets. While most observed free-floating planets are giant planets, we predict that there should be more low-mass free floating planets that are as yet unobserved than higher mass planets.
\end{abstract}



\section{Introduction}

Gravitational microlensing observations suggest that there could be a significant population of free-floating planets (FFPs) \citep[e.g.][]{Sumi2011,Clanton2017}. Estimates for the mass of FFPs range from around $0.25$ \citep{Mroz2017} up to about $3.5$ \citep{Sumi2011} Jupiter masses per main-sequence star. There is an excess of FFPs by a factor of up to seven compared to that predicted by core-collapse models \citep{Padoan2002,Miret-Roig2021}. There are several mechanisms suggested to form the excess of FFPs. These include planet-planet scattering \citep{Rasio1996,Weidenschilling1996,Veras2012}, aborted stellar embryo ejection from a stellar nursery \citep{Reipurth2001} and photo-erosion of a pre-stellar core  by stellar winds from a nearly OB star \citep{Whitworth2004}. Planet-planet scattering around single stars cannot explain the large number of FFPs \citep{Veras2012, Ma2016}.

Around a single star, planet-planet scattering occurs when the planets form close to each other. Two planets with masses $m_{\rm p1}$ and $m_{\rm p2}$ that form with semi-major axes $a_{\rm p1}$ and $a_{\rm p2}$, respectively, around a star with mass $m$ are unstable if  $\Delta=(a_{\rm p2}-a_{\rm p1})/R_{\rm Hill} \lesssim 2\sqrt{3}$  
\citep{Marchal1982,Gladman1993,Chambers1996}, where the  mutual Hill sphere radius is given by
\begin{equation}
R_{\rm Hill} = \left(\frac{m_{\rm p1}+m_{\rm p2}}{3 \,m}\right)^{1/3}\left(\frac{a_{\rm p1}+a_{\rm p2}}{2} \right).
\end{equation}
Ejection of giant planets in simulations of planet formation around a single star is unlikely and most ejected planets have a mass $\lesssim 0.3\,\rm M_\oplus$ \citep{Barclay2017}.
This stability criterion is not significantly affected if a single star is replaced by a {\it coplanar} inner binary, unless the planets are formed very close to the binary \citep{Kratter2014}.
However, the outcome of an unstable system is more likely to be ejection rather than  a collision around a binary star  system because of close encounters with the binary \citep{Smullen2016, Sutherland2016,Gong20171,Gong20172,Fleming2018}. Coplanar planets around a binary must form relatively close to each other for the system to be unstable.

Observations of circumbinary disks suggest that misalignment between the binary orbital plane and the disk may be common \citep[e.g.][]{Chiang2004,Winn2004,Capelo2012,Brinch2016, Zhu2022}. Polar aligned disks around eccentric binaries may also be common \citep{Kennedy2012,Kennedy2019,Kenworthy2022}. Chaotic accretion can lead to the formation of misaligned circumbinary disks \citep{clarke1993,Bate2018} and subsequent disk evolution can lead to tilt evolution towards a coplanar alignment \citep{PT1995, Bateetal2000,Lubow2000,Nixonetal2011a} or a polar alignment \citep{Aly2015,Martin2017,Lubow2018,Zanazzi2017}. However, the alignment timescale for an extended disk may be longer than the disk lifetime and planetary systems may form in a misaligned disk. While observations of circumbinary planets (CBPs) currently show only coplanar systems \citep[e.g.][]{Welsh2012,Orosz2012a,Kostov2013,Standing2023}, misaligned planetary systems may be expected around binaries with a longer orbital period \citep{Czekala2019}. These planetary systems may be observed in the future with eclipse-timing variations \citep{Zhang2019,MartinD2019}.

 Contrary to coplanar circumbinary planetary systems, a multi-planet system with a tilted respect to inner binary can be unstable for a wide range of initial planet separations \citep{Chen2023a,Chen2023b}. A planet orbiting around an eccentric binary  undergoes tilt oscillations as a result of its interaction with the binary \citep{Verrier2009,Farago2010,Naoz2017,Chen20192}. Planets in a tilted two-CBP system also undergo mutual tilt oscillations as a result of planet-planet interactions \citep{Chen2022}.  Furthermore, if the mutual tilt between the two planets becomes large, the outer planet can induce von-Zeipel-Kozai-Lidov \citep[ZKL][]{vonZeipel1910,Kozai1962,Lidov1962} oscillations of the inner planet that lead to planet eccentricity growth. Mean motion resonances (MMRs) between the planets are typically stable in a coplanar configuration, but they  become unstable in a tilted planetary system.  Therefore the complex dynamics around a tilted binary can lead to the planetary system instability for a wide range of the initial planet semi-major axis \citep{Chen2023a}. 


In this Letter, we consider the stability of tilted planetary systems with unequal mass planets. We show that the planet with the highest mass dominates the stability outcome.  The  order of the orbital radii of the planets does not change the stability; ejection of the smaller mass planets is the most likely outcome. We examine the planet mass required for planet-planet interactions to drive instability.  We describe our simulation setup and parameter domains we explore in Sec.~\ref{sim}. Specifically we consider the masses of the ejected planets.  Finally, a discussion and conclusions follow in Sec.~\ref{dis} and Sec.~\ref{con}.

\label{int}

\section{Simulation set--up and parameter space explored}
\label{sta}


\begin{table}
\centering
\caption{Parameters of the simulations. The first column contains the name of model, the second, third and fourth columns indicate the binary eccentricity and the masses of inner and outer planets. 
}
\begin{tabular}{cccc} 
\hline
\textbf{Model} & $e_{\rm b}$ & $m_{\rm p1}$ ($m_{\rm b}$) & $m_{\rm p2}$ ($m_{\rm b}$)  \\
\hline
\hline
A1 & 0.0 & 0.000003 & 0.001  \\
A2 & 0.8 & 0.000003 & 0.001   \\
C1 & 0.0 & 0.001 & 0.000003  \\
C2 & 0.8 & 0.001 & 0.000003  \\
\hline
E1 & 0.0 & 0.000003 & 0.000003  \\
E2 & 0.8 & 0.000003 & 0.000003 \\
SE1 & 0.0 & 0.00003 & 0.00003  \\
SE2 & 0.8 & 0.00003 & 0.00003  \\
N1 & 0.0 & 0.00005 & 0.00005  \\
N2 & 0.8 & 0.00005 & 0.00005  \\
S1 & 0.0 & 0.0003  & 0.0003  \\
S2 & 0.8 & 0.0003  & 0.0003  \\
\hline
\label{table1}
\end{tabular}
\end{table}

To study the orbital stability of 
planetary systems with unequal mass planets
we carry out simulations with the ${\sc n}$-body package, {\sc rebound} with a {\sc whfast} integrator which is a second order symplectic Wisdom Holman integrator with 11th order symplectic correctors \citep{Rein2015b}. We solve the gravitational equations in the frame of the center of mass of the four-body system. The central binary has components of mass $m_1$ and $m_2$ with the total mass $m_1+m_2=m_{\rm b}$. We consider an equal mass binary with $m_1=m_2=0.5 \,m_{\rm b}$ so that the mass fraction of the binary $f_{\rm b}$ = 0.5. The binary can be in a circular or eccentric orbit and the binary eccentricity is $e_{\rm b}$ and their separation is 1 $a_{\rm b}$. 

We consider planetary systems with two planets. We explored the extension to three planet systems in \cite{Chen2023b} where we found that the instability in the three planet case is qualitatively similar to the two planet case when the three planets are separated by a fixed number of mutual Hill radii. Three planet systems are slightly more unstable than two planet systems. The number of planets that are ejected in unstable systems depends upon how close the system is to the binary. Close-in systems are more likely to eject more planets. The planetary system arrangements for each simulation are shown in Table~\ref{table1}.

The two planets are initially on Keplerian orbits around the center of mass of the binary. 
Their orbits are defined by six orbital elements: the semi-major axes $a_{\rm p1}$ and $a_{\rm p2}$, inclinations $i_{\rm p1}$ and $i_{\rm p2}$ which is relative to the binary orbital plane, eccentricities $e_{\rm p1}$, and $e_{\rm p2}$, longitudes of the ascending nodes $\phi_{\rm p1}$ and $\phi_{\rm p2}$ measured from the binary semi--major axis, arguments of periapsides $\omega_{\rm p1}$, and $\omega_{\rm p2}$, and true anomalies $\nu_{\rm p1}$ and $\nu_{\rm p2}$. The initial orbits of two planets are coplanar to each other and circular so initially $e_{\rm p}=0$, $\omega_{\rm p}=0$ and $\nu_{\rm p}=0$ and we set $\phi_{\rm p}=90^{\circ}$ for all planets. We integrate the simulations for a total time of 14 million binary orbital periods ($T_{\rm b}$)  with a timestep of integration of 0.7$\%$ of the initial orbital period of the inner planet.

We describe the criteria for determining an unstable orbit of the planet based on three distinct conditions. First, if the eccentricity of the planet's orbit $e_{\rm p}\geq$ 1.0, the planet is deemed unbound from the binary system. Secondly, an orbit is  unstable if the semi-major axis of the planet's orbit $a_{\rm p}> 1000 \,a_{\rm b}$, indicating that the planet has moved excessively far away from the central system. Thirdly, if the semi-major axis of the planet's orbit becomes smaller than that of the binary's orbit, $a_{\rm p} < a_{\rm b}$, the planet can no longer be considered a CBP. These stability criteria are established in line with existing research \citep[e.g.][]{Chen20201, Quarles2020, Chen2023a, Chen2023b}.

To understand the dynamic interaction between two planets, the inner planet is placed at $a_{\rm p1}=5\,a_{\rm b}$ where it should be stable for a single CBP \citep{Chen20201}. We vary the semi-major axis of the outer planet by varying the ratio of their mutual Hill radius between the inner planet, $\Delta$ and thus, we have
\begin{equation}
    a_{\rm p2} = a_{\rm p1} + \Delta \left(\frac{m_{\rm p1}+m_{\rm p2}}{3 \,m_{\rm b}}\right)^{1/3}\left(\frac{a_{\rm p1}+a_{\rm p2}}{2} \right). 
\end{equation}
and $\Delta$ ranges from 3.4 $\left(\approx 2\sqrt{3} \right)$ to 12.0 with an interval of $\Delta = 0.1$. 

\section{Simulation results}
\label{sim}

In this Section we describe the results of our simulations. We first consider the effect of having an unequal mass planetary system, then the order of the planets in separation from the binary and finally the planet mass required for instability.

\subsection{The effect of unequal mass planetary systems} 

Fig.~\ref{fig:inner} shows stability maps in which the $x$-axis is the initial inclination of the planets with respect to the binary with an interval of 2.5$^{\circ}$ and the $y$-axis is the initial separation between two planets in units of $R_{\rm Hill}$. The colors of the pixels show the planets' orbital status at the end of simulations. The blue pixels represent systems in which the two planets are stable. The red pixels represent systems in which only the outer planet is stable and the green pixels  represent systems in which only the inner planet is stable. White pixels represent systems in which the two planets are unstable. The cyan, orange and pink horizontal dashed lines indicate the 2:1, 5:2 and 3:1 MMRs between the two planets. Note that the location of the MMRs are shifted relative to $\Delta$ compared to the equal Jupiter-mass planet cases in \citet{Chen2023b}. The total mass of the two CBPs in this study is smaller than two Jupiter-mass CBPs, so their separation in terms of $R_{\rm Hill}$ is smaller for a fixed $\Delta$.

The upper panels of Fig.~\ref{fig:inner} show  stability maps for models A1 and A2 that are planetary systems with an inner Earth-mass planet and an outer Jupiter mass planet with  binary eccentricity  $e_{\rm b}$ = 0.0 (left) and 0.8 (right). In model A1 with a circular orbit binary, if the two planets are near aligned to the binary orbital plane (in prograde or retrograde orbits), they can be stable even if they have vary small separation ($\Delta \geq 3.5$). However, with higher inclination orbits, the two planets are unlikely to be stable until $\Delta \geq 5.0$.  Above this region, there are unstable regions around the 2:1 and 3:1 MMRs  The effect of the 5:2 MMR is minor and there are sporadic unstable orbits around it. The only unstable outcome for a system with an inner Earth-mass and an outer Jupiter-mass planet is that the Earth mass planet gets ejected.

For model A2 with an eccentric binary, there are more inner planet unstable orbits within $\Delta \leq 5.0$ showing that misaligned planets are not likely  to be stable within the 2:1 MMR region except around the polar orbit ($i_{\rm p}$ around 90$^{\circ}$). All of the unstable cases are the outer planet stable cases meaning that the inner Earth-mass planet is ejected. Moreover, the unstable regions around 2:1 and 3:1 MMRs become larger than those of in model A1. Two planets are unlikely to be stable even when they are just slightly misaligned to the binary. We also increased the mass of the inner planet to one Saturn-mass and found most of the unstable cases again involve the ejection of the lower mass inner planet.

These two panels can be compared with with upper-left and lower-right panels of Figure~2 in \cite{Chen2023a} which show the same maps but for two Jupiter-mass CBPs. The units of the $y$-axis are different, but otherwise the stability map is very similar except that in the unequal mass system shown here, only the outer (and more massive) planet survives in all unstable cases. The two Jupiter-mass CBPs systems have a small fraction of systems in which both planets are ejected, however, with the Earth mass planet here, there are no systems in which both planets become ejected. Given the similarity between stability maps with two Jupiter mass planets and one Jupiter and one Earth, this suggests that the mass of the most massive planet in the system is what determines the stability of the system. 

\subsection{The effect of the planetary system order}

We now change the orbital arrangement of the two planets. The lower panels of Fig.~\ref{fig:inner} show models C1 and C2 that have the same setup as models A1 and A2 except the inner planet is the Jupiter-mass planet and the outer planet is the Earth-mass planet.  The stability map is not significantly affected by changing the order of the planets except that the outcome is different. The lower mass planet is ejected no matter where it begins and the Jupiter mass planet remains stable. The system is slightly more stable with the Jupiter mass planet being the inner planet. This is because an Earth mass planet that is interior to the more massive outer planet is more likely to undergo ZKL oscillations, resulting in increased eccentricity, interaction with the binary, and ejection when it is closer to the binary.



\subsection{The minimum mass to drive the instability}

Now we consider the minimum mass of the planet required to drive the planet-planet interactions for instability. If the effect of planet-planet interactions is small, each CBP only interacts with the binary so than their orbital stability should comply with the three-body stability map \citep[e.g.][]{Doolin2011, Quarles2018, Chen20201}. We have shown that the most massive planet in the system dominates the stability already so we now consider systems of equal masses but we vary the mass.
We consider two CBPs of Earth-mass (models E1 and E2), 10 Earth-mass (models SE1 and SE2), Neptune-mass (models N1 and N2) and Saturn-mass (models S1 and S2). 

\begin{figure*}
  \centering
    \includegraphics[width=7.5cm]{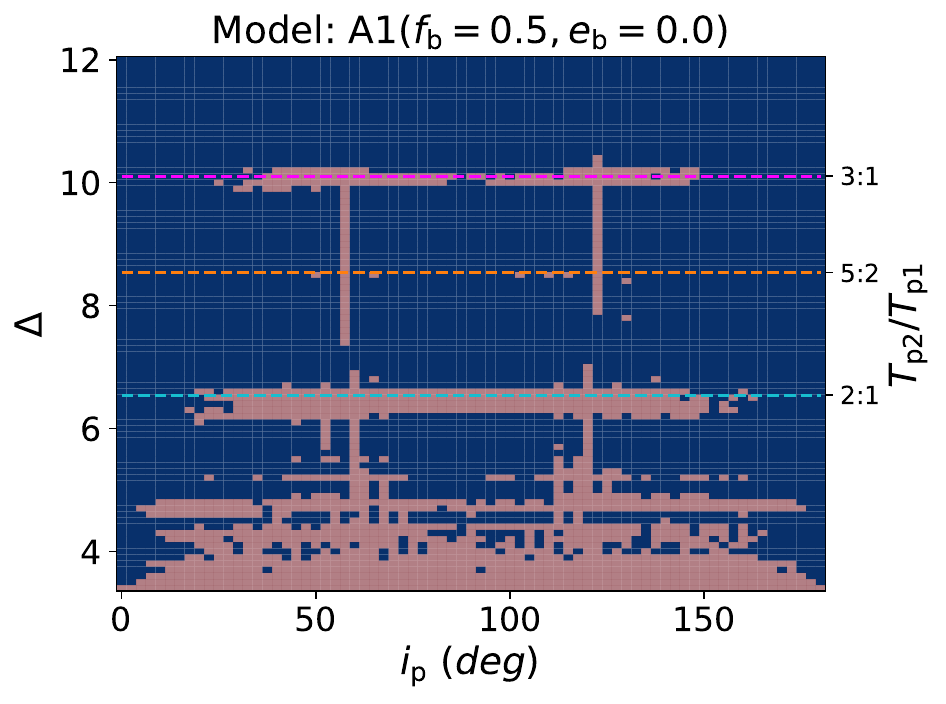}
    \includegraphics[width=7.5cm]{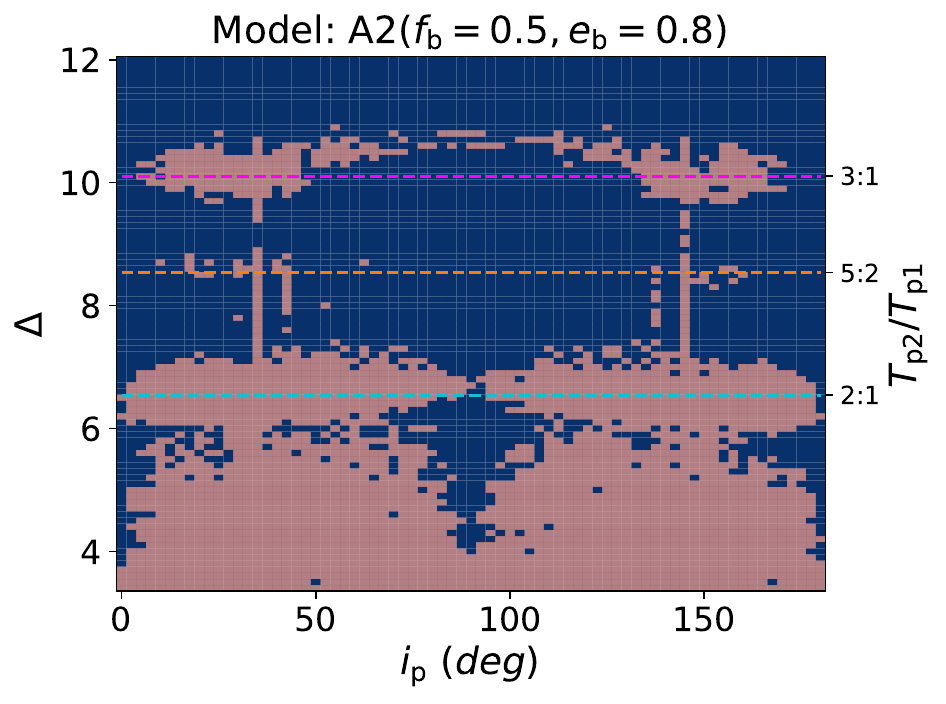}
    \includegraphics[width=7.5cm]{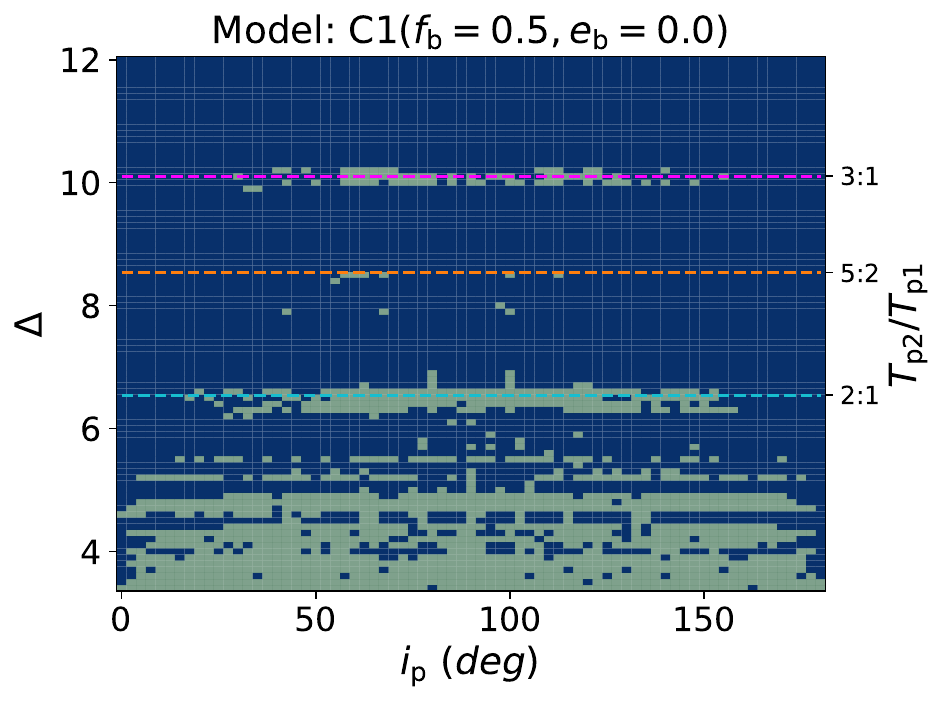}
    \includegraphics[width=7.5cm]{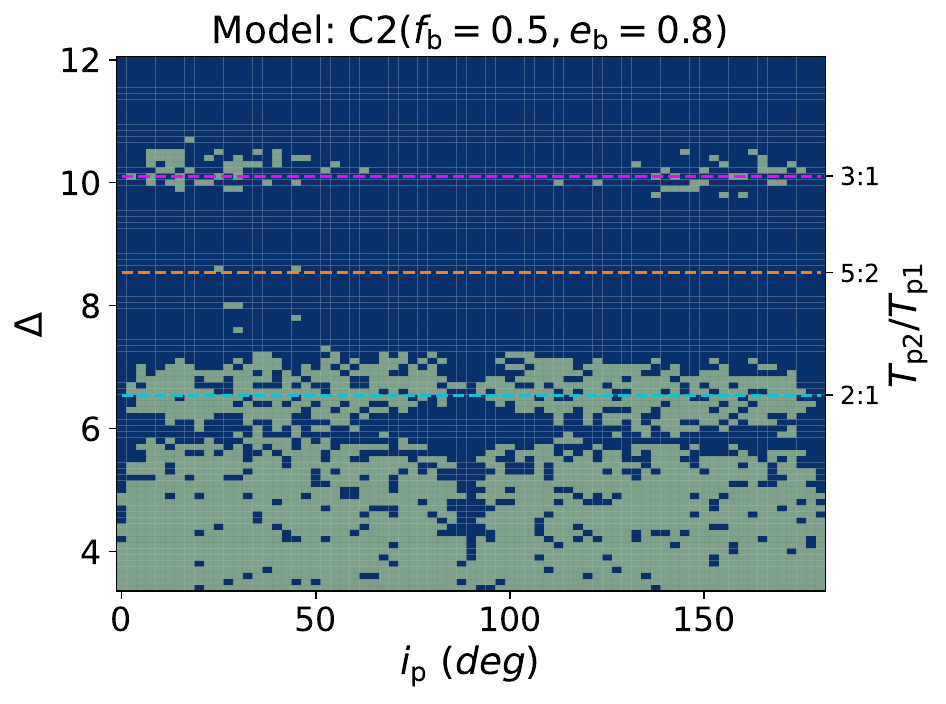}
    
    \caption{Stability maps of a two planet circumbinary system with $e_{\rm b} =0.0$ (left panels), 0.8 (right panels). The planetary system consists of  one Earth mass planet and one Jupiter mass planet. The Jupiter-mass planet is the outer planet on upper panels and is the inner planet on lower panels. The inner planet has initial semi-major axis $a_{\rm p1} = 5\, a_{\rm b}$ and the $y$-axis is $\Delta$ with an interval of $0.1$ while the $x$-axis is the initial inclination of two planets with an interval of $2.5^{\circ}$. The four horizontal dashed lines represent the 2:1 (cyan), 5:2 (orange), 3:1 (pink) and 4:1 (red) MMRs between the two planets. The four different colors of pixels represent two planets stable cases (blue), the inner planet survived cases (green), the outer planet survived cases (red) and both planet unstable cases (white). } 
     \label{fig:inner}
\end{figure*}

\begin{figure*}
  \centering
    \includegraphics[width=7.5cm]{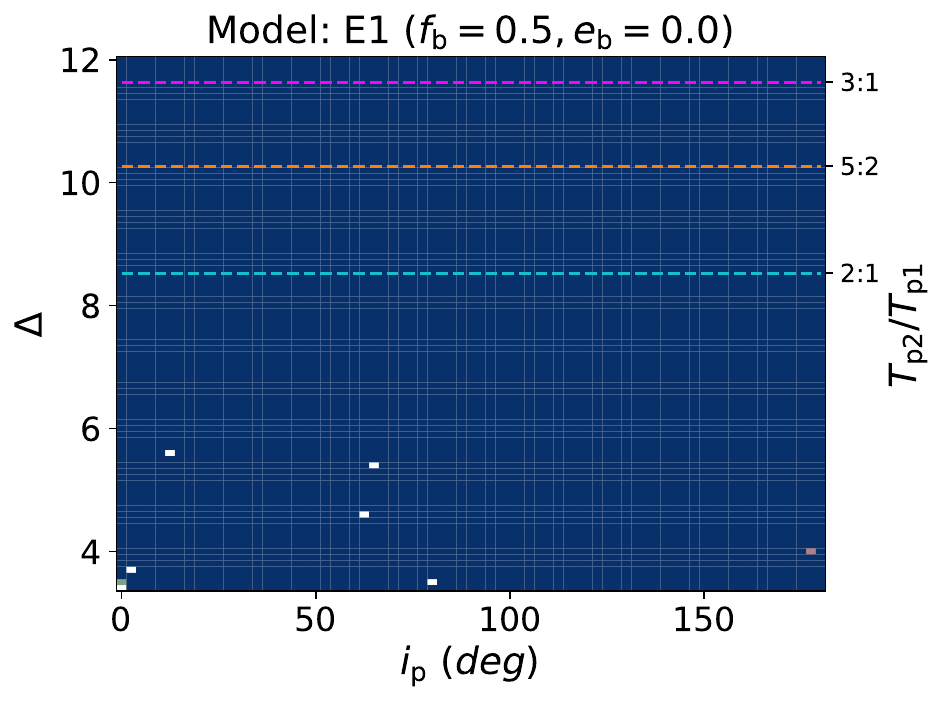}
    \includegraphics[width=7.5cm]{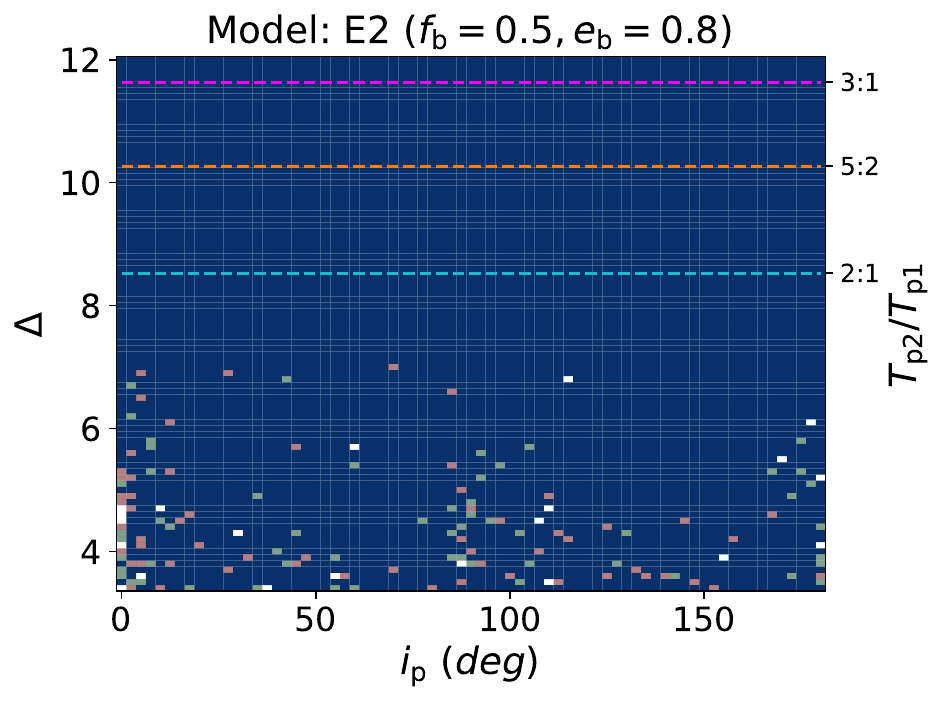}
    \includegraphics[width=7.5cm]{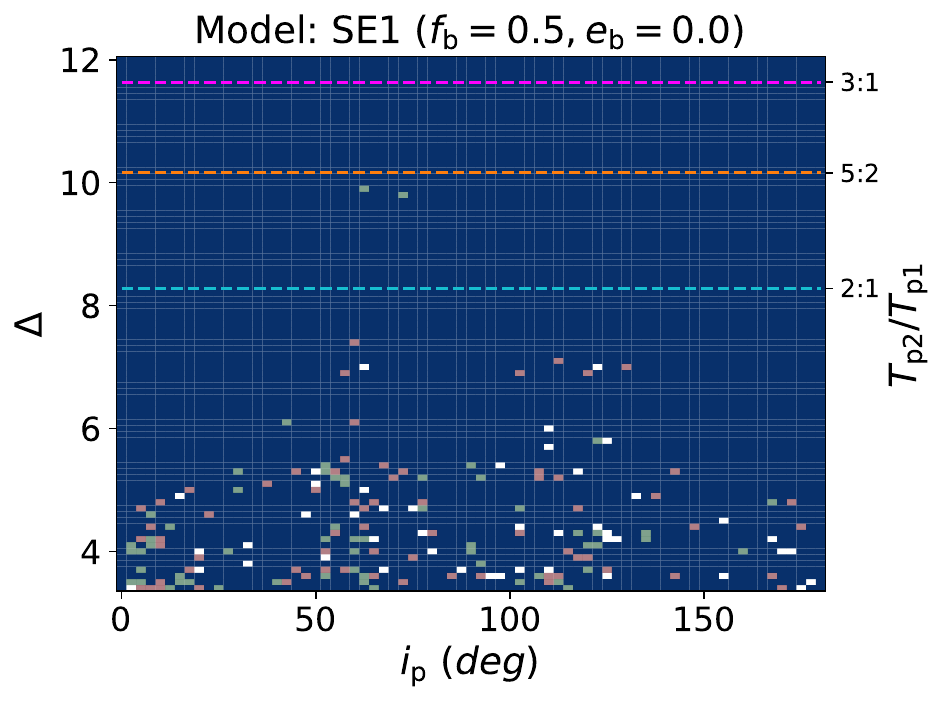}
    \includegraphics[width=7.5cm]{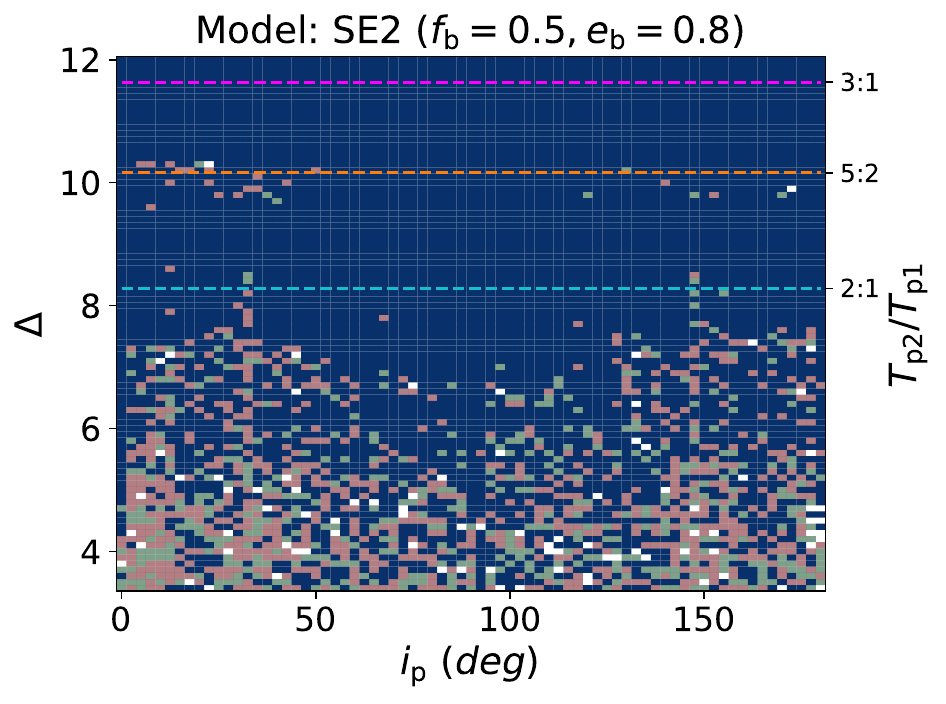}
    \includegraphics[width=7.5cm]{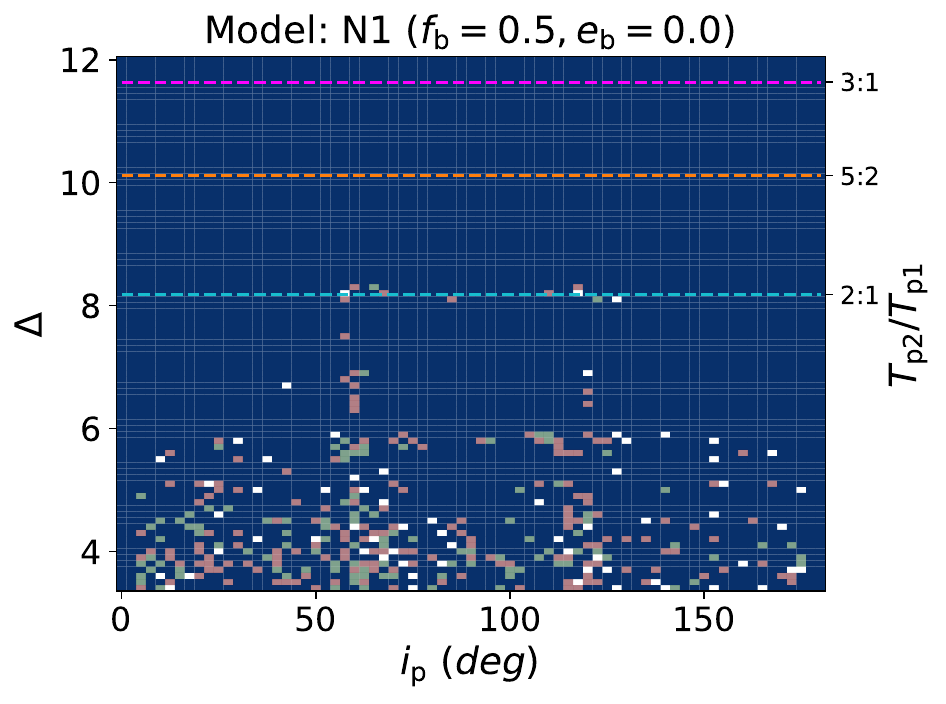}
    \includegraphics[width=7.5cm]{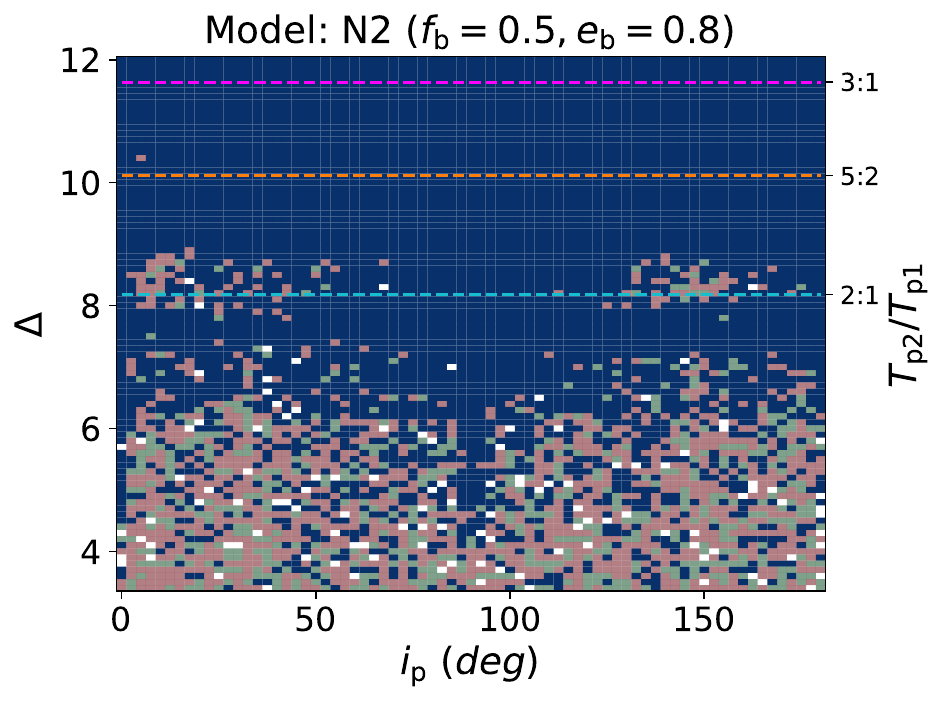}
    \includegraphics[width=7.5cm]{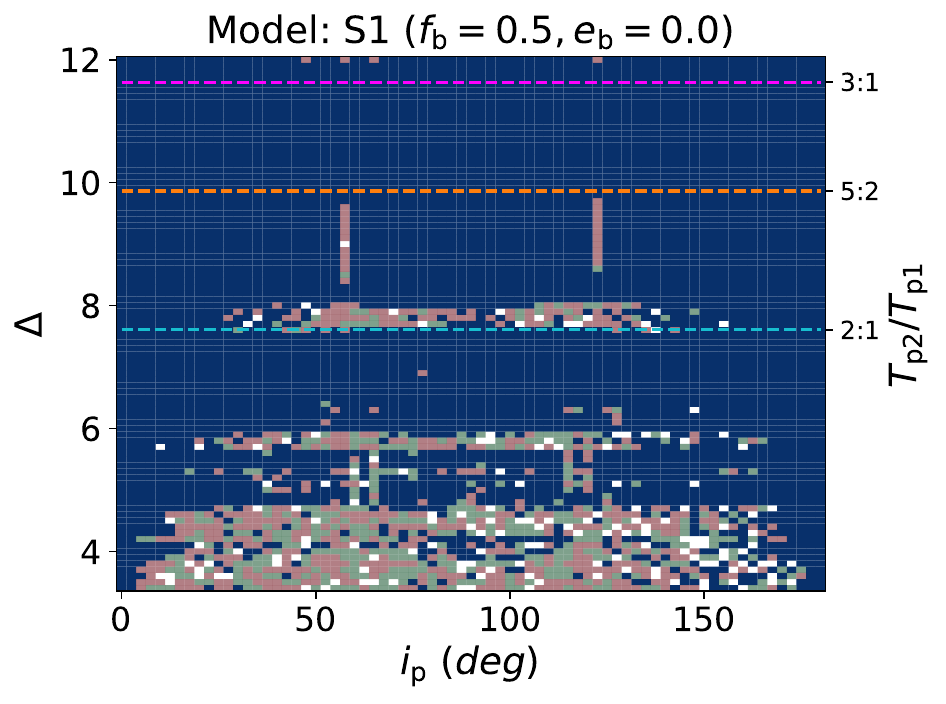}
    \includegraphics[width=7.5cm]{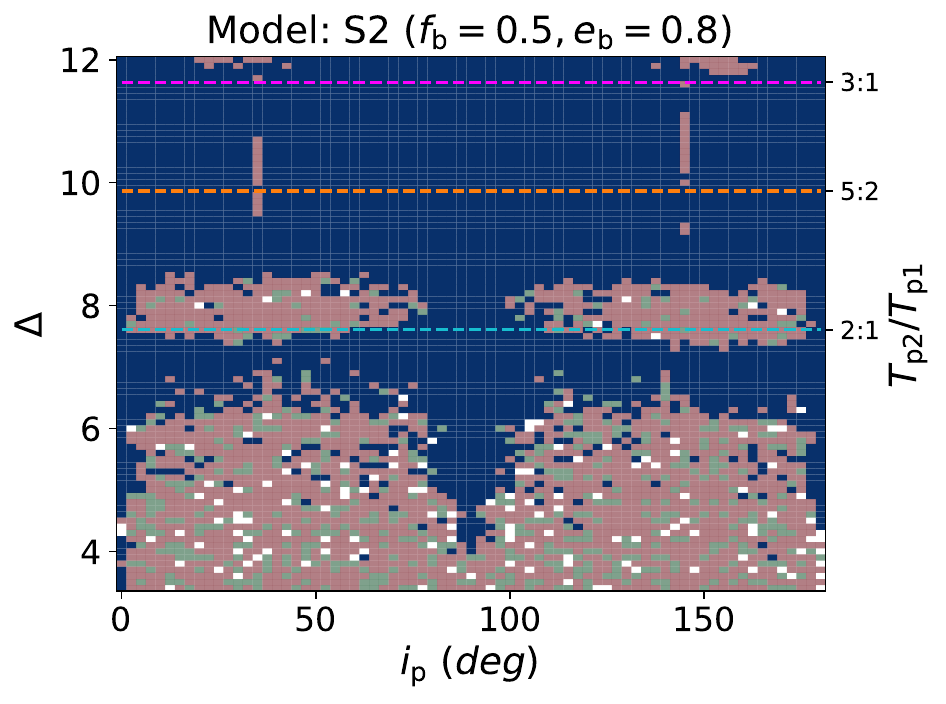}
    
    \caption{Stability maps for planetary systems with two equal-mass circumbinary planets around a binary system with $e_{\rm b} =0.0$ (left panels), 0.8 (right panels). The inner planet has initial semi-major axis $a_{\rm p1} = 5\, a_{\rm b}$. The pixel colors are as described in Fig.~\ref{fig:inner}. From upper to lower panels, the two planets are Earth-mass, Super-Earth mass (10 times Earth-mass), Neptune-mass and Saturn-mass.} 
     \label{fig:mass}
\end{figure*}

The left  panels of Fig.~\ref{fig:mass} show the effect of increasing the planet masses around a circular orbit binary. Two Earth-mass CBPs (model E1) are very stable. With increasing masses of two planets, the number of unstable cases increases. For model S1 in which the planets are Saturn-mass, the stability map is very similar to the map of A1 in Fig.~\ref{fig:inner} with Jupiter mass planets. Around a circular orbit binary, instability requires a planet mass greater than about Neptune's mass to drive planetary system instability.

The right panels of Fig.~\ref{fig:mass} show the same simulations with increasing planet masses but around an eccentric orbit binary 
 with $e_{\rm b} = 0.8$. The Earth mass planets  are still stable but there some unstable cases at separations closer than than the 2:1 MMR.  For the super-Earth mass planets, more than half of the pixels are unstable for separations closer  than the 2:1 MMR in models SE2 and N2. For model S2 which planets are Saturn-mass, the map is very similar to map of A2 in Fig.~\ref{fig:inner} and two planets can be only stable in near coplanar or polar orbits within the 2:1 MMR. Around a highly eccentric orbit binary, instability requires smaller mass planets than around circular orbit binaries. Planet-planet interactions drive instability for planet masses greater than super-Earth masses for these cases which have $a_{\rm p1} = 5 a_{\rm b}$.

\section{Discussion}
\label{dis}

Planet-planet scattering, gravitational instabilities in the outer disk and post-formation evolution around single star systems may contribute some population of the observed FFPs, but it is difficult to explain the large populations that we have observed  \citep{Boss2011, Veras2012}. Recently, some N-body simulations have confirmed this point. For example, simulations with Jupiter and Saturn mass planets around a solar-type Star in \citet{Barclay2017} show that approximately one-third of the initial disk mass is ejected ($\sim$ 5 Earth-mass) in the late stage of planet formation but the masses of the individual ejected bodies are $\lesssim 0.3$ Earth-mass. Simulations in \citet{Pfyffer2015} found that the median mass of an ejected planet is $\leq$ 2 Earth-mass from post-formation evolution. We also investigated multi-planetary single star systems with two Jupiter-mass planets  and found that they are very stable unless they are extremely compactly spaced with $\Delta \leq 2.3$  \citep[see Figure 1 of][]{Chen2023a}.

On the other hand, binary stars are common in the Universe, and multiplicity statistics show that about 40$\%$ to 60$\%$ of M to F stars are in binary systems. Approximately half of these binaries have separations of less than 20 au, indicating that they may have CBDs orbiting around them during the star formation \citep[see Figure 12 and 13 in][]{Raghavan2010}. Since CBDs have a wide distribution of inclinations if the binary period $T_{\rm b}$ exceeds if $T_{\rm b} > 30$ days \citep{Czekala2019} due to the broad distribution of binary eccentricities $e_{\rm b}$ \citep[see Figure 14 in][]{Raghavan2010}, planets could form inside these misaligned disks. Furthermore, the detection of disk substructures such as rings and gaps in the HL Tau disk \citep{ALMA2015,Yen2016, Yen2019} implies that early planet formation can occur in disks \citep[cf.][]{Nixon2018}. Several planet-induced structures can be explained by numerical simulations, although the topic is still under debate \citep[e.g.][]{Dong2015, Flock2015, Zhang2015, Zhang2018}. Hence, planet formation could be very efficient, and we expect that multiple planets could also form in the CBD around the binary.

If planet formation occurs while the disk is young, it is likely that multiple planets form within the disk that can either decouple from the disk \citep[e.g.][]{Martin2016,franchini2020} and be ejected or be ejected once the disk has dispersed \citep[cf.][]{Turrini2020}. Recent observations show that 93$\%$ of exoplanet pairs are at least 10$\,R_{\rm Hill}$ apart \citep{Weiss2018} and CBPs pairs may have a similar distribution due to instabilities within the 3:1 MMR. A binary with a lower eccentricity has a higher chance to host a compact CBP pair while only a compact Earth-mass CBP pair could be stable around a binary with high eccentricity. Consequently, FFPs ranging from low to high mass can be generated from the circumbinary planetary systems. 

In all of our simulations we have chosen the inner planet semi-major axis to be $a_{\rm p1} = 5.0 a_{\rm b}$ since this is close to the inner stability limit for inclined orbits of all inclinations. According to Figure 3 and 5 in \citet{Chen2023a}, the instabilities are stronger for misaligned CBPs if the inner  planet is at 10.0 or 20.0 $a_{\rm b}$ due to the ZKL-like oscillations. 
However, if the inner planet is much farther from the binary, then the system behaves as though the binary is a single star. The only instability is then due to planet-planet interactions and the system becomes more stable.

From Transiting Exoplanet Survey Satellite (TESS) data, the giant planet (0.6 -- 2.0$R_{\rm J}$) occurrence rate is about 0.194$\pm$ 0.072 per cent for low mass stars (0.088--0.7$M_{\odot}$) \citep{Bryant2023} and about 7.8 percent for giant stars from the Pan-Pacific Planet Search \citep{Wittenmyer2020}. On the other hand, the occurrence rate of small planets ranging from (0.5--4.0 $R_{\oplus}$) with orbital periods shorter than 50 days is about 90 percent from the Kepler Input Catalog \citep{Dressing2013}. If these results are similar to the planet occurrence rates in circumbinary systems, there should be a large number of low mass CBPs which we have not detected yet or they have been ejected out from their original systems. Furthermore, even though all confirmed CBPs are nearly coplanar to their host binary orbits, recent observations show that the range of the gas disk inclinations around the binary can be wide \citep{Czekala2019}. Therefore, CBPs that are formed inside a misaligned disks may be destabilized due to massive CBPs after the disk is dispersed.

Very recently, the James Webb Space Telescope (JWST) near-infrared survey of the inner Orion Nebula and Trapezium Cluster has found that 9$\%$ of the planetary mass objects ($\geq$ 0.6 Jupiter masses) are in wide pairs of planets with similar mass objects \citep{Pearson2023}. These sub-stellar binaries may form from the same circumbinary planetary systems where both planets ejected. However, it is not at all clear that pair formation is possible.  If pairs do form, their mass ratios would not be extreme. 


\section{Conclusions}
\label{con}

In this study, we considered the ejection of inclined orbit interacting circumbinary planets as a possible origin of free-floating planets. We explored the orbital stability of two planet systems around a binary with equal mass stars each of half Solar mass. We considered cases of two interacting CBPs with an initial inner planet orbital radius $a_{\rm p1} = 5 a_{\rm b}$. At this inner orbital radius, the planet is stable to binary-planet interactions for a system with a single planet \citep{Chen20201} The planets are initially mutually coplanar, but inclined relative to the binary orbital plane.
 
We first considered cases with unequal mass planets including a Jupiter-mass and an Earth-mass planet. Comparing with previous stability maps with two Jupiter-mass planets \citep[Figure 2 of][]{Chen2023a}, we find that the stability maps with different mass planets are slightly more stable, but qualitatively very similar. However, in this letter we find that only the more massive Jupiter-mass planet can survive in the unstable cases. In addition, initial misalignment of the planets plays an important role in providing instability.  Thus, a Jupiter-mass planet tends to kick out other low mass CBPs if they are in initially misaligned orbits with respect to the binary. A low mass CBP can be slightly more stable if it is located outside of the massive CBP.

We examined the dependence of the orbital instability on planet mass by considering CBP systems with equal mass planets for several values of planet mass and two values of binary eccentricity. We find that instability generally increases with increasing planet mass and with increasing binary eccentricity. 

The results show that CBPs with $m_{\rm p}\gtrsim$ Neptune-mass around a circular or eccentric orbit binary are unstable, while CBPs with each planet having mass $m_{\rm p}\gtrsim$ 10 Earth-mass are unstable around an eccentric binary. Two CBPs with $m_{\rm p}\geq$ Saturn-mass have a similar stability as two Jupiter-mass planets. Consequently, we may find more stable compact planetary systems around low $e_{\rm b}$ binaries. Notice that our simulations are scale free so that the minimum mass to trigger this instability can vary with the total mass of the binary. 

In conclusion, we predict there are significantly more small mass planets compared to high mass planets that are ejected  from binary star systems. This may be in agreement with the result that shows that there is a 95\% upper limit on the frequency of free-floating Jupiter mass planets that is 0.25 per star \citep{Mroz2017}. In addition, recent observations for the Milky Way and M31 by the Subaru Hyper Suprime-Cam show that there are abundant populations of  unbound sub-terrestrial objects with masses in range of $10^{-5} M_{\oplus}-1 M_{\oplus}$ \citep{DeRocco2023}. Considering that low mass planets are more numerous and the natural instability of a circumbinary system with a massive CBP, we suggest that binary systems can generate both high and low mass FFPs due to the combined effects of planet-planet and binary-planet interactions. 
 
 Apart from TESS, in an upcoming observation season, the Nancy Grace Roman Space Telescope (Roman) survey will explore the Galactic bulge for FFPs \citep{Penny2019, Johnson2020}. Furthermore, the capabilities of the Roman mission extend to the prospect of detecting FFPs within the Magellanic Clouds, as suggested by recent research in \citet{Sajadian2021}. Finally, JWST has detected many free-floating binary planets. We will investigate further whether the CBP model has the potential to address the origin of these objects. 

\begin{acknowledgements}

Computer support was provided by UNLV's National Supercomputing Center and the DiRAC Data Intensive service at Leicester, operated by the University of Leicester IT Services, which forms part of the STFC DiRAC HPC Facility (www.dirac.ac.uk). CC and CJN acknowledge support from the Science and Technology Facilities Council (grant number ST/Y000544/1). CC thanks for the useful discussion with Dr. Wei Zhu on the PPVII conference. RGM and SHL acknowledge support from NASA through grants 80NSSC19K0443 and 80NSSC21K0395. CJN acknowledges support from the Leverhulme Trust (grant number RPG-2021-380).  Simulations in this letter made use of the REBOUND code which can be downloaded freely at http://github.com/hannorein/rebound.
\end{acknowledgements}

\bibliography{main}

\begin{thebibliography}{}
\expandafter\ifx\csname natexlab\endcsname\relax\def\natexlab#1{#1}\fi
\providecommand{\url}[1]{\href{#1}{#1}}
\providecommand{\dodoi}[1]{doi:~\href{http://doi.org/#1}{\nolinkurl{#1}}}
\providecommand{\doeprint}[1]{\href{http://ascl.net/#1}{\nolinkurl{http://ascl.net/#1}}}
\providecommand{\doarXiv}[1]{\href{https://arxiv.org/abs/#1}{\nolinkurl{https://arxiv.org/abs/#1}}}

\bibitem[{{ALMA Partnership} {et~al.}(2015){ALMA Partnership}, {Brogan}, {P{\'e}rez}, {Hunter}, {Dent}, {Hales}, {Hills}, {Corder}, {Fomalont}, {Vlahakis}, {Asaki}, {Barkats}, {Hirota}, {Hodge}, {Impellizzeri}, {Kneissl}, {Liuzzo}, {Lucas}, {Marcelino}, {Matsushita}, {Nakanishi}, {Phillips}, {Richards}, {Toledo}, {Aladro}, {Broguiere}, {Cortes}, {Cortes}, {Espada}, {Galarza}, {Garcia-Appadoo}, {Guzman-Ramirez}, {Humphreys}, {Jung}, {Kameno}, {Laing}, {Leon}, {Marconi}, {Mignano}, {Nikolic}, {Nyman}, {Radiszcz}, {Remijan}, {Rod{\'o}n}, {Sawada}, {Takahashi}, {Tilanus}, {Vila Vilaro}, {Watson}, {Wiklind}, {Akiyama}, {Chapillon}, {de Gregorio-Monsalvo}, {Di Francesco}, {Gueth}, {Kawamura}, {Lee}, {Nguyen Luong}, {Mangum}, {Pietu}, {Sanhueza}, {Saigo}, {Takakuwa}, {Ubach}, {van Kempen}, {Wootten}, {Castro-Carrizo}, {Francke}, {Gallardo}, {Garcia}, {Gonzalez}, {Hill}, {Kaminski}, {Kurono}, {Liu}, {Lopez}, {Morales}, {Plarre}, {Schieven}, {Testi}, {Videla}, {Villard}, {Andreani}, {Hibbard}, \&
  {Tatematsu}}]{ALMA2015}
{ALMA Partnership}, {Brogan}, C.~L., {P{\'e}rez}, L.~M., {et~al.} 2015, \apjl, 808, L3, \dodoi{10.1088/2041-8205/808/1/L3}

\bibitem[{{Aly} {et~al.}(2015){Aly}, {Dehnen}, {Nixon}, \& {King}}]{Aly2015}
{Aly}, H., {Dehnen}, W., {Nixon}, C., \& {King}, A. 2015, MNRAS, 449, 65, \dodoi{10.1093/mnras/stv128}

\bibitem[{{Barclay} {et~al.}(2017){Barclay}, {Quintana}, {Raymond}, \& {Penny}}]{Barclay2017}
{Barclay}, T., {Quintana}, E.~V., {Raymond}, S.~N., \& {Penny}, M.~T. 2017, \apj, 841, 86, \dodoi{10.3847/1538-4357/aa705b}

\bibitem[{Bate(2018)}]{Bate2018}
Bate, M.~R. 2018, \mnras, 475, 5618, \dodoi{10.1093/mnras/sty169}

\bibitem[{{Bate} {et~al.}(2000){Bate}, {Bonnell}, {Clarke}, {Lubow}, {Ogilvie}, {Pringle}, \& {Tout}}]{Bateetal2000}
{Bate}, M.~R., {Bonnell}, I.~A., {Clarke}, C.~J., {et~al.} 2000, MNRAS, 317, 773, \dodoi{10.1046/j.1365-8711.2000.03648.x}

\bibitem[{{Boss}(2011)}]{Boss2011}
{Boss}, A.~P. 2011, \apj, 731, 74, \dodoi{10.1088/0004-637X/731/1/74}

\bibitem[{{Brinch} {et~al.}(2016){Brinch}, {J{\o}rgensen}, {Hogerheijde}, {Nelson}, \& {Gressel}}]{Brinch2016}
{Brinch}, C., {J{\o}rgensen}, J.~K., {Hogerheijde}, M.~R., {Nelson}, R.~P., \& {Gressel}, O. 2016, \apjl, 830, L16, \dodoi{10.3847/2041-8205/830/1/L16}

\bibitem[{Bryant {et~al.}(2023)Bryant, Bayliss, \& Van~Eylen}]{Bryant2023}
Bryant, E.~M., Bayliss, D., \& Van~Eylen, V. 2023, Monthly Notices of the Royal Astronomical Society, 521, 3663

\bibitem[{{Capelo} {et~al.}(2012){Capelo}, {Herbst}, {Leggett}, {Hamilton}, \& {Johnson}}]{Capelo2012}
{Capelo}, H.~L., {Herbst}, W., {Leggett}, S.~K., {Hamilton}, C.~M., \& {Johnson}, J.~A. 2012, \apjl, 757, L18, \dodoi{10.1088/2041-8205/757/1/L18}

\bibitem[{Chambers {et~al.}(1996)Chambers, Wetherill, \& Boss}]{Chambers1996}
Chambers, J., Wetherill, G., \& Boss, A. 1996, Icarus, 119, 261, \dodoi{https://doi.org/10.1006/icar.1996.0019}

\bibitem[{Chen {et~al.}(2019)Chen, Franchini, Lubow, \& Martin}]{Chen20192}
Chen, C., Franchini, A., Lubow, S.~H., \& Martin, R.~G. 2019, \mnras, 490, 5634, \dodoi{10.1093/mnras/stz2948}

\bibitem[{Chen {et~al.}(2020)Chen, Lubow, \& Martin}]{Chen20201}
Chen, C., Lubow, S.~H., \& Martin, R.~G. 2020, \mnras, 494, 4645, \dodoi{10.1093/mnras/staa1037}

\bibitem[{{Chen} {et~al.}(2022){Chen}, {Lubow}, \& {Martin}}]{Chen2022}
{Chen}, C., {Lubow}, S.~H., \& {Martin}, R.~G. 2022, \mnras, 510, 351, \dodoi{10.1093/mnras/stab3488}

\bibitem[{{Chen} {et~al.}(2023){Chen}, {Lubow}, {Martin}, \& {Nixon}}]{Chen2023a}
{Chen}, C., {Lubow}, S.~H., {Martin}, R.~G., \& {Nixon}, C.~J. 2023, \mnras, 521, 5033, \dodoi{10.1093/mnras/stad739}

\bibitem[{Chen {et~al.}(2023)Chen, Martin, \& Nixon}]{Chen2023b}
Chen, C., Martin, R.~G., \& Nixon, C.~J. 2023, Monthly Notices of the Royal Astronomical Society, 525, 3781, \dodoi{10.1093/mnras/stad2543}

\bibitem[{{Chiang} \& {Murray-Clay}(2004)}]{Chiang2004}
{Chiang}, E.~I., \& {Murray-Clay}, R.~A. 2004, \apj, 607, 913, \dodoi{10.1086/383522}

\bibitem[{{Clanton} \& {Gaudi}(2017)}]{Clanton2017}
{Clanton}, C., \& {Gaudi}, B.~S. 2017, \apj, 834, 46, \dodoi{10.3847/1538-4357/834/1/46}

\bibitem[{{Clarke} \& {Pringle}(1993)}]{clarke1993}
{Clarke}, C.~J., \& {Pringle}, J.~E. 1993, \mnras, 261, 190, \dodoi{10.1093/mnras/261.1.190}

\bibitem[{Czekala {et~al.}(2019)Czekala, Chiang, Andrews, Jensen, Torres, Wilner, Stassun, \& Macintosh}]{Czekala2019}
Czekala, I., Chiang, E., Andrews, S.~M., {et~al.} 2019, \apj, 883, 22, \dodoi{10.3847/1538-4357/ab287b}

\bibitem[{{DeRocco} {et~al.}(2023){DeRocco}, {Smyth}, \& {Profumo}}]{DeRocco2023}
{DeRocco}, W., {Smyth}, N., \& {Profumo}, S. 2023, arXiv e-prints, arXiv:2308.13593, \dodoi{10.48550/arXiv.2308.13593}

\bibitem[{{Dong} {et~al.}(2015){Dong}, {Zhu}, \& {Whitney}}]{Dong2015}
{Dong}, R., {Zhu}, Z., \& {Whitney}, B. 2015, \apj, 809, 93, \dodoi{10.1088/0004-637X/809/1/93}

\bibitem[{{Doolin} \& {Blundell}(2011)}]{Doolin2011}
{Doolin}, S., \& {Blundell}, K.~M. 2011, \mnras, 418, 2656, \dodoi{10.1111/j.1365-2966.2011.19657.x}

\bibitem[{Dressing \& Charbonneau(2013)}]{Dressing2013}
Dressing, C.~D., \& Charbonneau, D. 2013, The Astrophysical Journal, 767, 95, \dodoi{10.1088/0004-637X/767/1/95}

\bibitem[{{Farago} \& {Laskar}(2010)}]{Farago2010}
{Farago}, F., \& {Laskar}, J. 2010, \mnras, 401, 1189, \dodoi{10.1111/j.1365-2966.2009.15711.x}

\bibitem[{{Fleming} {et~al.}(2018){Fleming}, {Barnes}, {Graham}, {Luger}, \& {Quinn}}]{Fleming2018}
{Fleming}, D.~P., {Barnes}, R., {Graham}, D.~E., {Luger}, R., \& {Quinn}, T.~R. 2018, \apj, 858, 86, \dodoi{10.3847/1538-4357/aabd38}

\bibitem[{{Flock} {et~al.}(2015){Flock}, {Ruge}, {Dzyurkevich}, {Henning}, {Klahr}, \& {Wolf}}]{Flock2015}
{Flock}, M., {Ruge}, J.~P., {Dzyurkevich}, N., {et~al.} 2015, \aap, 574, A68, \dodoi{10.1051/0004-6361/201424693}

\bibitem[{{Franchini} {et~al.}(2020){Franchini}, {Martin}, \& {Lubow}}]{franchini2020}
{Franchini}, A., {Martin}, R.~G., \& {Lubow}, S.~H. 2020, \mnras, 491, 5351, \dodoi{10.1093/mnras/stz3175}

\bibitem[{{Gladman}(1993)}]{Gladman1993}
{Gladman}, B. 1993, \icarus, 106, 247, \dodoi{10.1006/icar.1993.1169}

\bibitem[{{Gong}(2017)}]{Gong20171}
{Gong}, Y.-X. 2017, \apj, 834, 55, \dodoi{10.3847/1538-4357/834/1/55}

\bibitem[{{Gong} \& {Ji}(2017)}]{Gong20172}
{Gong}, Y.-X., \& {Ji}, J. 2017, \aj, 154, 179, \dodoi{10.3847/1538-3881/aa8c7c}

\bibitem[{Johnson {et~al.}(2020)Johnson, Penny, Gaudi, Kerins, Rattenbury, Robin, Novati, \& Henderson}]{Johnson2020}
Johnson, S.~A., Penny, M., Gaudi, B.~S., {et~al.} 2020, \aj, 160, 123, \dodoi{10.3847/1538-3881/aba75b}

\bibitem[{{Kennedy} {et~al.}(2012){Kennedy}, {Wyatt}, {Sibthorpe}, {Duch{\^e}ne}, {Kalas}, {Matthews}, {Greaves}, {Su}, \& {Fitzgerald}}]{Kennedy2012}
{Kennedy}, G.~M., {Wyatt}, M.~C., {Sibthorpe}, B., {et~al.} 2012, \mnras, 421, 2264, \dodoi{10.1111/j.1365-2966.2012.20448.x}

\bibitem[{{Kennedy} {et~al.}(2019){Kennedy}, {Matr{\`a}}, {Facchini}, {Milli}, {Pani{\'c}}, {Price}, {Wilner}, {Wyatt}, \& {Yelverton}}]{Kennedy2019}
{Kennedy}, G.~M., {Matr{\`a}}, L., {Facchini}, S., {et~al.} 2019, Nature Astronomy, 3, 278, \dodoi{10.1038/s41550-019-0715-1}

\bibitem[{{Kenworthy} {et~al.}(2022){Kenworthy}, {Gonz{\'a}lez Picos}, {Elizondo}, {Martin}, {van Dam}, {Rodriguez}, {Kennedy}, {Ginski}, {Mugrauer}, {Vogt}, {Adam}, \& {Oelkers}}]{Kenworthy2022}
{Kenworthy}, M.~A., {Gonz{\'a}lez Picos}, D., {Elizondo}, E., {et~al.} 2022, \aap, 666, A61, \dodoi{10.1051/0004-6361/202243441}

\bibitem[{{Kostov} {et~al.}(2013){Kostov}, {McCullough}, {Hinse}, {Tsvetanov}, {H{\'e}brard}, {D{\'\i}az}, {Deleuil}, \& {Valenti}}]{Kostov2013}
{Kostov}, V.~B., {McCullough}, P.~R., {Hinse}, T.~C., {et~al.} 2013, \apj, 770, 52, \dodoi{10.1088/0004-637X/770/1/52}

\bibitem[{{Kozai}(1962)}]{Kozai1962}
{Kozai}, Y. 1962, \aj, 67, 591, \dodoi{10.1086/108790}

\bibitem[{{Kratter} \& {Shannon}(2014)}]{Kratter2014}
{Kratter}, K.~M., \& {Shannon}, A. 2014, \mnras, 437, 3727, \dodoi{10.1093/mnras/stt2179}

\bibitem[{{Lidov}(1962)}]{Lidov1962}
{Lidov}, M.~L. 1962, \planss, 9, 719, \dodoi{10.1016/0032-0633(62)90129-0}

\bibitem[{{Lubow} \& {Martin}(2018)}]{Lubow2018}
{Lubow}, S.~H., \& {Martin}, R.~G. 2018, \mnras, 473, 3733, \dodoi{10.1093/mnras/stx2643}

\bibitem[{{Lubow} \& {Ogilvie}(2000)}]{Lubow2000}
{Lubow}, S.~H., \& {Ogilvie}, G.~I. 2000, \apj, 538, 326, \dodoi{10.1086/309101}

\bibitem[{{Ma} {et~al.}(2016){Ma}, {Mao}, {Ida}, {Zhu}, \& {Lin}}]{Ma2016}
{Ma}, S., {Mao}, S., {Ida}, S., {Zhu}, W., \& {Lin}, D. N.~C. 2016, \mnras, 461, L107, \dodoi{10.1093/mnrasl/slw110}

\bibitem[{{Marchal} \& {Bozis}(1982)}]{Marchal1982}
{Marchal}, C., \& {Bozis}, G. 1982, Celestial Mechanics, 26, 311, \dodoi{10.1007/BF01230725}

\bibitem[{Martin(2019)}]{MartinD2019}
Martin, D.~V. 2019, Monthly Notices of the Royal Astronomical Society, 488, 3482, \dodoi{10.1093/mnras/stz959}

\bibitem[{{Martin} \& {Lubow}(2017)}]{Martin2017}
{Martin}, R.~G., \& {Lubow}, S.~H. 2017, \apjl, 835, L28, \dodoi{10.3847/2041-8213/835/2/L28}

\bibitem[{{Martin} {et~al.}(2016){Martin}, {Lubow}, {Nixon}, \& {Armitage}}]{Martin2016}
{Martin}, R.~G., {Lubow}, S.~H., {Nixon}, C., \& {Armitage}, P.~J. 2016, \mnras, 458, 4345, \dodoi{10.1093/mnras/stw605}

\bibitem[{Miret-Roig {et~al.}(2021)Miret-Roig, Bouy, Raymond, Tamura, Bertin, Barrado, Olivares, Galli, Cuillandre, Sarro, \& et~al.}]{Miret-Roig2021}
Miret-Roig, N., Bouy, H., Raymond, S.~N., {et~al.} 2021, Nature Astronomy, 6, 89–97, \dodoi{10.1038/s41550-021-01513-x}

\bibitem[{{Mr{\'o}z} {et~al.}(2017){Mr{\'o}z}, {Udalski}, {Skowron}, {Poleski}, {Koz{\l}owski}, {Szyma{\'n}ski}, {Soszy{\'n}ski}, {Wyrzykowski}, {Pietrukowicz}, {Ulaczyk}, {Skowron}, \& {Pawlak}}]{Mroz2017}
{Mr{\'o}z}, P., {Udalski}, A., {Skowron}, J., {et~al.} 2017, \nat, 548, 183, \dodoi{10.1038/nature23276}

\bibitem[{{Naoz} {et~al.}(2017){Naoz}, {Li}, {Zanardi}, {de El{\'\i}a}, \& {Di Sisto}}]{Naoz2017}
{Naoz}, S., {Li}, G., {Zanardi}, M., {de El{\'\i}a}, G.~C., \& {Di Sisto}, R.~P. 2017, \aj, 154, 18, \dodoi{10.3847/1538-3881/aa6fb0}

\bibitem[{{Nixon} {et~al.}(2011){Nixon}, {Cossins}, {King}, \& {Pringle}}]{Nixonetal2011a}
{Nixon}, C.~J., {Cossins}, P.~J., {King}, A.~R., \& {Pringle}, J.~E. 2011, MNRAS, 412, 1591, \dodoi{10.1111/j.1365-2966.2010.17952.x}

\bibitem[{{Nixon} {et~al.}(2018){Nixon}, {King}, \& {Pringle}}]{Nixon2018}
{Nixon}, C.~J., {King}, A.~R., \& {Pringle}, J.~E. 2018, \mnras, 477, 3273, \dodoi{10.1093/mnras/sty593}

\bibitem[{{Orosz} {et~al.}(2012){Orosz}, {Welsh}, {Carter}, {Brugamyer}, {Buchhave}, {Cochran}, {Endl}, {Ford}, {MacQueen}, {Short}, {Torres}, {Windmiller}, {Agol}, {Barclay}, {Caldwell}, {Clarke}, {Doyle}, {Fabrycky}, {Geary}, {Haghighipour}, {Holman}, {Ibrahim}, {Jenkins}, {Kinemuchi}, {Li}, {Lissauer}, {Pr{\v{s}}a}, {Ragozzine}, {Shporer}, {Still}, \& {Wade}}]{Orosz2012a}
{Orosz}, J.~A., {Welsh}, W.~F., {Carter}, J.~A., {et~al.} 2012, \apj, 758, 87, \dodoi{10.1088/0004-637X/758/2/87}

\bibitem[{{Padoan} \& {Nordlund}(2002)}]{Padoan2002}
{Padoan}, P., \& {Nordlund}, {\r{A}}. 2002, \apj, 576, 870, \dodoi{10.1086/341790}

\bibitem[{{Papaloizou} \& {Terquem}(1995)}]{PT1995}
{Papaloizou}, J.~C.~B., \& {Terquem}, C. 1995, MNRAS, 274, 987

\bibitem[{Pearson \& McCaughrean(2023)}]{Pearson2023}
Pearson, S.~G., \& McCaughrean, M.~J. 2023, Jupiter Mass Binary Objects in the Trapezium Cluster.
\newblock \doarXiv{2310.01231}

\bibitem[{{Penny} {et~al.}(2019){Penny}, {Gaudi}, {Kerins}, {Rattenbury}, {Mao}, {Robin}, \& {Calchi Novati}}]{Penny2019}
{Penny}, M.~T., {Gaudi}, B.~S., {Kerins}, E., {et~al.} 2019, \apj, 241, 3, \dodoi{10.3847/1538-4365/aafb69}

\bibitem[{{Pfyffer} {et~al.}(2015){Pfyffer}, {Alibert}, {Benz}, \& {Swoboda}}]{Pfyffer2015}
{Pfyffer}, S., {Alibert}, Y., {Benz}, W., \& {Swoboda}, D. 2015, \aap, 579, A37, \dodoi{10.1051/0004-6361/201424295}

\bibitem[{Quarles {et~al.}(2020)Quarles, Li, Kostov, \& Haghighipour}]{Quarles2020}
Quarles, B., Li, G., Kostov, V., \& Haghighipour, N. 2020, The Astronomical Journal, 159, 80, \dodoi{10.3847/1538-3881/ab64fa}

\bibitem[{{Quarles} {et~al.}(2018){Quarles}, {Satyal}, {Kostov}, {Kaib}, \& {Haghighipour}}]{Quarles2018}
{Quarles}, B., {Satyal}, S., {Kostov}, V., {Kaib}, N., \& {Haghighipour}, N. 2018, \apj, 856, 150, \dodoi{10.3847/1538-4357/aab264}

\bibitem[{{Raghavan} {et~al.}(2010){Raghavan}, {McAlister}, {Henry}, {Latham}, {Marcy}, {Mason}, {Gies}, {White}, \& {ten Brummelaar}}]{Raghavan2010}
{Raghavan}, D., {McAlister}, H.~A., {Henry}, T.~J., {et~al.} 2010, \apjs, 190, 1, \dodoi{10.1088/0067-0049/190/1/1}

\bibitem[{{Rasio} \& {Ford}(1996)}]{Rasio1996}
{Rasio}, F.~A., \& {Ford}, E.~B. 1996, Science, 274, 954, \dodoi{10.1126/science.274.5289.954}

\bibitem[{{Rein} \& {Tamayo}(2015)}]{Rein2015b}
{Rein}, H., \& {Tamayo}, D. 2015, \mnras, 452, 376, \dodoi{10.1093/mnras/stv1257}

\bibitem[{{Reipurth} \& {Clarke}(2001)}]{Reipurth2001}
{Reipurth}, B., \& {Clarke}, C. 2001, \aj, 122, 432, \dodoi{10.1086/321121}

\bibitem[{{Sajadian}(2021)}]{Sajadian2021}
{Sajadian}, S. 2021, \mnras, 506, 3615, \dodoi{10.1093/mnras/stab1907}

\bibitem[{{Smullen} {et~al.}(2016){Smullen}, {Kratter}, \& {Shannon}}]{Smullen2016}
{Smullen}, R.~A., {Kratter}, K.~M., \& {Shannon}, A. 2016, \mnras, 461, 1288, \dodoi{10.1093/mnras/stw1347}

\bibitem[{{Standing} {et~al.}(2023){Standing}, {Sairam}, {Martin}, {Triaud}, {Correia}, {Coleman}, {Baycroft}, {Kunovac}, {Boisse}, {Collier Cameron}, {Dransfield}, {Faria}, {Gillon}, {Hara}, {Hellier}, {Howard}, {Lane}, {Mardling}, {Maxted}, {Miller}, {Nelson}, {Orosz}, {Pepe}, {Santerne}, {Sebastian}, {Udry}, \& {Welsh}}]{Standing2023}
{Standing}, M.~R., {Sairam}, L., {Martin}, D.~V., {et~al.} 2023, arXiv e-prints, arXiv:2301.10794, \dodoi{10.48550/arXiv.2301.10794}

\bibitem[{{Sumi} {et~al.}(2011){Sumi}, {Kamiya}, {Bennett}, {Bond}, {Abe}, {Botzler}, {Fukui}, {Furusawa}, {Hearnshaw}, {Itow}, {Kilmartin}, {Korpela}, {Lin}, {Ling}, {Masuda}, {Matsubara}, {Miyake}, {Motomura}, {Muraki}, {Nagaya}, {Nakamura}, {Ohnishi}, {Okumura}, {Perrott}, {Rattenbury}, {Saito}, {Sako}, {Sullivan}, {Sweatman}, {Tristram}, {Udalski}, {Szyma{\'n}ski}, {Kubiak}, {Pietrzy{\'n}ski}, {Poleski}, {Soszy{\'n}ski}, {Wyrzykowski}, {Ulaczyk}, \& {Microlensing Observations in Astrophysics (MOA) Collaboration}}]{Sumi2011}
{Sumi}, T., {Kamiya}, K., {Bennett}, D.~P., {et~al.} 2011, \nat, 473, 349, \dodoi{10.1038/nature10092}

\bibitem[{{Sutherland} \& {Fabrycky}(2016)}]{Sutherland2016}
{Sutherland}, A.~P., \& {Fabrycky}, D.~C. 2016, \apj, 818, 6, \dodoi{10.3847/0004-637X/818/1/6}

\bibitem[{{Turrini} {et~al.}(2020){Turrini}, {Zinzi}, \& {Belinchon}}]{Turrini2020}
{Turrini}, D., {Zinzi}, A., \& {Belinchon}, J.~A. 2020, \aap, 636, A53, \dodoi{10.1051/0004-6361/201936301}

\bibitem[{{Veras} \& {Raymond}(2012)}]{Veras2012}
{Veras}, D., \& {Raymond}, S.~N. 2012, \mnras, 421, L117, \dodoi{10.1111/j.1745-3933.2012.01218.x}

\bibitem[{{Verrier} \& {Evans}(2009)}]{Verrier2009}
{Verrier}, P.~E., \& {Evans}, N.~W. 2009, \mnras, 394, 1721, \dodoi{10.1111/j.1365-2966.2009.14446.x}

\bibitem[{{von Zeipel}(1910)}]{vonZeipel1910}
{von Zeipel}, H. 1910, Astronomische Nachrichten, 183, 345, \dodoi{10.1002/asna.19091832202}

\bibitem[{{Weidenschilling} \& {Marzari}(1996)}]{Weidenschilling1996}
{Weidenschilling}, S.~J., \& {Marzari}, F. 1996, \nat, 384, 619, \dodoi{10.1038/384619a0}

\bibitem[{{Weiss} {et~al.}(2018){Weiss}, {Marcy}, {Petigura}, {Fulton}, {Howard}, {Winn}, {Isaacson}, {Morton}, {Hirsch}, {Sinukoff}, {Cumming}, {Hebb}, \& {Cargile}}]{Weiss2018}
{Weiss}, L.~M., {Marcy}, G.~W., {Petigura}, E.~A., {et~al.} 2018, \aj, 155, 48, \dodoi{10.3847/1538-3881/aa9ff6}

\bibitem[{{Welsh} {et~al.}(2012){Welsh}, {Orosz}, {Carter}, {Fabrycky}, {Ford}, {Lissauer}, {Pr{\v s}a}, {Quinn}, \& {et al.}}]{Welsh2012}
{Welsh}, W.~F., {Orosz}, J.~A., {Carter}, J.~A., {et~al.} 2012, \nat, 481, 475, \dodoi{10.1038/nature10768}

\bibitem[{{Whitworth} \& {Zinnecker}(2004)}]{Whitworth2004}
{Whitworth}, A.~P., \& {Zinnecker}, H. 2004, \aap, 427, 299, \dodoi{10.1051/0004-6361:20041131}

\bibitem[{{Winn} {et~al.}(2004){Winn}, {Holman}, {Johnson}, {Stanek}, \& {Garnavich}}]{Winn2004}
{Winn}, J.~N., {Holman}, M.~J., {Johnson}, J.~A., {Stanek}, K.~Z., \& {Garnavich}, P.~M. 2004, \apjl, 603, L45, \dodoi{10.1086/383089}

\bibitem[{{Wittenmyer} {et~al.}(2020){Wittenmyer}, {Butler}, {Horner}, {Clark}, {Tinney}, {Carter}, {Wang}, {Johnson}, \& {Collins}}]{Wittenmyer2020}
{Wittenmyer}, R.~A., {Butler}, R.~P., {Horner}, J., {et~al.} 2020, \mnras, 491, 5248, \dodoi{10.1093/mnras/stz3378}

\bibitem[{{Yen} {et~al.}(2019){Yen}, {Gu}, {Hirano}, {Koch}, {Lee}, {Liu}, \& {Takakuwa}}]{Yen2019}
{Yen}, H.-W., {Gu}, P.-G., {Hirano}, N., {et~al.} 2019, \apj, 880, 69, \dodoi{10.3847/1538-4357/ab29f8}

\bibitem[{{Yen} {et~al.}(2016){Yen}, {Liu}, {Gu}, {Hirano}, {Lee}, {Puspitaningrum}, \& {Takakuwa}}]{Yen2016}
{Yen}, H.-W., {Liu}, H.~B., {Gu}, P.-G., {et~al.} 2016, \apjl, 820, L25, \dodoi{10.3847/2041-8205/820/2/L25}

\bibitem[{{Zanazzi} \& {Lai}(2017)}]{Zanazzi2017}
{Zanazzi}, J.~J., \& {Lai}, D. 2017, \mnras, 467, 1957, \dodoi{10.1093/mnras/stx208}

\bibitem[{{Zhang} {et~al.}(2015){Zhang}, {Blake}, \& {Bergin}}]{Zhang2015}
{Zhang}, K., {Blake}, G.~A., \& {Bergin}, E.~A. 2015, \apjl, 806, L7, \dodoi{10.1088/2041-8205/806/1/L7}

\bibitem[{{Zhang} {et~al.}(2018){Zhang}, {Zhu}, {Huang}, {Guzm{\'a}n}, {Andrews}, {Birnstiel}, {Dullemond}, {Carpenter}, {Isella}, {P{\'e}rez}, {Benisty}, {Wilner}, {Baruteau}, {Bai}, \& {Ricci}}]{Zhang2018}
{Zhang}, S., {Zhu}, Z., {Huang}, J., {et~al.} 2018, \apjl, 869, L47, \dodoi{10.3847/2041-8213/aaf744}

\bibitem[{{Zhang} \& {Fabrycky}(2019)}]{Zhang2019}
{Zhang}, Z., \& {Fabrycky}, D.~C. 2019, \apj, 879, 92, \dodoi{10.3847/1538-4357/ab24d5}

\bibitem[{{Zhu} {et~al.}(2022){Zhu}, {Bernhard}, {Dai}, {Fang}, {Zanazzi}, {Zang}, {Dong}, {Hambsch}, {Gan}, {Wu}, \& {Poon}}]{Zhu2022}
{Zhu}, W., {Bernhard}, K., {Dai}, F., {et~al.} 2022, \apjl, 933, L21, \dodoi{10.3847/2041-8213/ac7b2d}

\end{thebibliography}
\bibliographystyle{aasjournal}

\end{document}